\documentclass[%
 reprint,
 amsmath,amssymb,
 aps,amsthm,
]{revtex4-2}
\usepackage{physics}
\usepackage{graphicx}
\usepackage{dcolumn}
\usepackage{bm}

\usepackage[utf8]{inputenc}
\usepackage{enumerate}
\usepackage{mathtools}
\usepackage{algorithm,algorithmic}
\usepackage{blindtext}
\usepackage{graphicx}
\graphicspath{ {./images/} }
\usepackage{epsfig}
\usepackage{sidecap}
\usepackage{hyperref,stackrel}
\usepackage[normalem]{ulem}
\usepackage{color}
\usepackage{enumerate}
\usepackage{bm}
\usepackage{amsthm,amsmath}
\usepackage{amsfonts}
\usepackage{amssymb,wasysym}
\usepackage{graphicx}
\usepackage{enumitem}
\usepackage{lmodern}
\usepackage{braket}
\usepackage[mathscr]{euscript}

\usepackage{multirow}
\usepackage [english]{babel}
\usepackage [autostyle, english = american]{csquotes}
\MakeOuterQuote{"}
\usepackage{caption}
\usepackage{subcaption}

\theoremstyle{definition}
\newtheorem{definition}{Definition}[section]
\newtheorem{prop}{Proposition}

\makeatletter
\newcommand*{\rom}[1]{\expandafter\@slowromancap\romannumeral #1@}
\makeatother
\begin{document}

\preprint{APS/123-QED}

\title{Entanglement Routing and Bottlenecks in Grid Networks}

\author{Vaisakh Mannalath}\email{vaisakhmannalath@gmail.com}

\author{Anirban Pathak}
 \email{anirban.pathak@jiit.ac.in}
\affiliation{%
 Department of Physics and Materials Science \& Engineering, Jaypee Institute of Information Technology, A-10, Sector 62, Noida, UP-201309, India
  }%

\date{\today}

                        
\begin{abstract}
Distributing entangled pairs among multiple users is a fundamental problem in quantum networks. Existing protocols like $X$ protocol introduced in (npj Quantum Information 5, 76 (2019)) use graph theoretic tools like local complementation to optimize the number of measurements required to extract any Bell pair among the network users. However, such a  protocol relies on finding the shortest path between the users. Here, the existing results are extended to establish a counter-intuitive notion that, in general, the most optimal path to perform the $X$ protocol is not along the shortest path. Specific examples of this advantage are provided on networks of size as small as 12 qubits. Bottlenecks in establishing simultaneous Bell pairs in nearest-neighbor architectures are also explored. Recent results suggesting the unsuitability of the line and ring networks for the implementation of quantum networks due to the existence of bottlenecks are revisited, and using local equivalency relations from graph theory, it is hinted at the possibility that even grid graphs are not exempt from bottleneck issues. Further, it's noted that the results obtained here would be of use in analyzing the advantages of measurement-based quantum network coding.
\end{abstract}

\maketitle


\section{Introduction}Point-to-point secure quantum communication over optical fibers and through free space has witnessed significant advancement over the years \cite{CZL+21,ACS+21}. Nevertheless, transmission errors and losses limit the distance of direct quantum communication \cite{PLO+17}, prompting the application of entanglement swapping protocols to address this challenge \cite{SSR+11,ATL15}. This has enabled a number of successful experiments over increasing distances and success rates \cite{YCZ+08,LZY+19,LTM+21}, such as the intercontinental quantum communication between China and Austria via a satellite \cite{LCH+18}. This has presented opportunities for further progress to be made in terms of maximum distance \cite{CZL+22} and communication rates \cite{CZC+21}.

However, the development of algorithms to manage the entanglement in a quantum internet and distribute entangled states between multiple nodes (users) is a fundamental and complex problem \cite{LLY+19,MMG19,BSK+20}. Graph states are useful tools in the study of quantum networks \cite{S03,HEB04}; they have been utilized to accomplish tasks such as quantum metrology \cite{SM20}, quantum error correcting codes \cite{SW01}, and one-way quantum computing \cite{RB01}. In addition, various schemes for generating entanglement among specific network nodes have been proposed \cite{LLC21,MBF10,PPE+22,PKT+19}, and some of them have been implemented experimentally \cite{HKM+18,KRH+17,ZCB+21,SNN+20}. Despite this, the optimal protocol for remote entanglement generation in a quantum network is yet to be determined. To this end, Ref. \cite{DHW20} investigated whether a given graph state can be transformed into a set of Bell states between specific network nodes with operations restricted to single-qubit Clifford operations, single-qubit Pauli measurements, and classical communication. It was concluded that this problem is \texttt{NP-Complete}, and thus, it is necessary to devise better-performing protocols for certain instances pertinent to multipartite schemes.

In this regard, a general method to extract maximally entangled states with two  parties in connected networks was outlined in \cite{HPE19}. Compared to other methods, such as the algorithm described in  \cite{SMI+16}, which requires large amounts of quantum memories, this approach was more advantageous regarding the memory required for the repeater stations. The protocol proposed in \cite{HPE19} is a repeater-based protocol involving manipulating an already generated graph state to accommodate future communication requests. This method is quite similar to a standard repeater-based protocol, albeit more efficient. In the graph state formalism, the maximally entangled states shared by two parties are represented by line graphs, with two  vertices, up to some local operations. One could perform sequential entanglement swapping measurements on a path connecting all two nodes to establish a maximally entangled state between two  nodes. This process consists of performing a series of entanglement swapping operations, which allow for the transfer of entanglement from one node to another. By exploiting the entanglement swapping operations, the protocol proposed in \cite{HPE19} allows for a much more efficient generation of maximally entangled states between two (three) nodes in connected networks.

 In \cite{HPE19}, a protocol for entanglement routing was proposed for a scenario where a graph state is already shared among the network users, with a single qubit of memory per user. The protocol was based on graph theoretic tools like local complementation. It demonstrates how it can be leveraged to reduce the number of measurements required compared to standard repeater-based schemes. It was also shown that local complementation can solve bottleneck issues in network architectures like the  \emph{butterfly network}. Here we aim to extend the protocol introduced in \cite{HPE19}, increasing its efficiency and the domain of applicability. Specifically, we show how the original proof can be modified to incorporate various scenarios and provide examples demonstrating the modified protocol's advantage. Furthermore, we provide a method to apply recent results on bottlenecks in ring graphs \cite{HDE+22} to illustrate how grid graphs are affected by it. Using this, we demonstrate how the butterfly and similar networks suffer from the same bottleneck issues as ring and line graphs.  \\

 The rest of the paper is organized as follows; in Section \ref{sec:prelim}, we briefly introduce graph states and graph theory tools used in this paper. 
Then, in Section \ref{sec:xproto}, we state and prove our result on Bell pair generation in a network. In Section \ref{sec:bottle}, we consider a class of butterfly-like networks and relate their bottleneck issues to that of the rising graphs. In Section \ref{sec:discussion} we conclude with our remarks and possible future research questions.

\section{Preliminaries}
\label{sec:prelim}
A finite graph $G=(V, E)$ is specified by a set of vertices $V \subsetneq \mathbb{N}$ and a set $E \subseteq V \times V$ of edges. A simple graph is one without any loop (an edge that links a vertex to itself) or multiple edges connecting the same pair of vertices. For any given vertex $a$, the set of all vertices connected by an edge is known as the neighborhood of $a$ and is symbolized as $N_a$.

\begin{definition}
 (Vertex Deletion): Deleting a vertex $v$ results in a graph where the vertex $v$ and all the edges connected to it are removed.
 $$G-v=\left(V \backslash v,\{e \in E: e \cap v=\varnothing\}\right)$$.
 \end{definition}
\begin{definition}
(Local complementation): A local complementation $LC_v$ is a graph operation specified by a vertex $v$, taking a graph $G$ to $LC_v(G)$ by replacing the neighborhood of $v$ by its complement.
$$
LC_v(G)=\left( V,E\Delta K_{N_v}\right),
$$
where $K_{N_v}$ is the set of edges of the complete graph
on the vertex set $N_v$ and $E \Delta K_{N_v}=(E \cup K_{N_v})-(E \cap K_{N_v})$ is the symmetric difference.
\end{definition}
Local complementation acts on the neighbourhood of a vertex by removing edges if they are present and
adding missing edges, if any.
\begin{definition}
(Vertex-minor): A graph $H$ is called a vertex-minor of $G$ if a sequence of local complementations and vertex-deletions maps $G$ to $H$.
\end{definition}

   The graph $G\left( V,E\right)$ can be associated with a pure quantum state, referred to as a graph state, which is defined on a Hilbert space $\mathcal{H}_V=\left(\mathbb{C}^{2}\right)^{\otimes V}$. Specifically, each vertex in $V$ is assigned a qubit in the state $|+\rangle=(|0\rangle+|1\rangle) / \sqrt{2}$. Then, a controlled-$Z$ operation is applied to any two qubits connected by an edge, resulting in the graph state $|G\rangle$ for the graph $G\left( V,E\right)$, as described in \cite{HEB04}.Thus, a graph state is defined as follows; 
$$|G\rangle:=\prod_{(i, j) \in E} C Z_{i, j}|+\rangle^{\otimes V}.$$

Local Clifford operations on graph states can be implemented through local complementations on the related graph \cite{VDD04}. Moreover, local Pauli measurements on graph states can be represented using the combination of local complementations and vertex deletions \cite{HEB04}. We can visualize the role played by the Pauli measurements as follows.

  \begin{prop}
 ($Z$-measurement) Measurement of a qubit, corresponding to the vertex $v$, in the $Z$-basis is represented by the vertex deletion of $v$.
 $$Z_v(G)=G-v
 $$
 \end{prop}
 \begin{prop}
 \label{prop:ymeas}
 ($Y$-measurement) Measurement of a qubit, corresponding to the vertex $v$, in the $Y$-basis is represented by,
 $$Y_v(G)=Z_vLC_v(G)
 $$
 \end{prop}
 \begin{prop}
 \label{prop:xmeas}
 ($X$-measurement) Measurement of a qubit, corresponding to the vertex $v$, in the $X$-basis is represented by
 $$X_v(G)=LC_wZ_vLC_vLC_w(G)
 $$
 where $w\in N_v$.
 \end{prop}

\section{X protocol}
\label{sec:xproto}
Given a network state, there is a straightforward way to extract a Bell pair shared among network users. It is similar to the repeater-based approach of sharing entanglement between two distant parties and is hence called the \emph{repeater protocol}. In repeater protocol, a path is first isolated between two given vertices $a$ and $b$ by performing $Z$ measurements on the neighboring vertices. The vertices $a$ and $b$ are then connected via the $X$ measurements of the intermediate vertices on the path. The total number of measurements can be optimized by choosing a path with the minimum combined neighborhood. The authors of \emph{Ref.} \cite{HPE19} introduced the \emph{$X$ protocol} where the $X$ measurement on the intermediate vertices along the path is carried out prior to the isolation step. The sequential $X$ measurements on the intermediate vertices connect $a$ and $b$. The two nodes are then disconnected from the rest of the graph via $Z$ measurements of the neighboring vertices. The authors proved that along the shortest path connecting two vertices $a,b$, the $X$ protocol requires at most as many measurements as the repeater protocol (cf. Theorem 1 of \cite{HPE19}).

An observation to be made at this point is that for the repeater protocol, it is enough that the path of choice should be one with the minimum combined neighborhood, which, when isolated, forms a \emph{repeater line}. We define the repeater line to be a path for which the set of edges $E(V, V)=\{(v_1,v_2),\cdots,(v_{n-1},v_{n})\}$ for vertices $V=\{v_1,\cdots,v_n\}$. It does not matter whether such a repeater line is the shortest or not since the total number of measurements (for the repeater protocol) depends solely on the cardinality of the combined neighborhood of the path, which we have taken to be a minimum. Naturally, the question arises whether the $X$ protocol holds the same advantage over the repeater protocol in this generalized case as well. We answer this in the positive and demonstrate examples where taking the "longer" path is more advantageous than the shorter one.

We first show how the $X$ protocol works in a repeater line. Let $v_1, v_n$ be the vertices to be connected. Consider $\{(v_1,v_2),\cdots,(v_{n-1},v_{n})\}$ to be a repeater line connecting them. Now, we may examine how these edges change with $X$ measurements along the repeater line. $X$ measurement on a vertex is equivalent to locally complementing a neighbor, then locally complementing the actual
node and deleting it, followed by final local complementation of the same neighbor Equation \ref{prop:xmeas}. For example, let us look at how $X$ measurement on $v_2$ updates the edge set of the repeater line. Since $v_2$ is connected to $v_1$, we choose it as a neighbor for the $X$ measurement. It is easy to see that measuring $v_2$ connects $v_1$ and $v_3$. We can proceed similarly to the $X$ measurement of $v_3$ and the rest of the vertices, which ultimately connect $v_1$ to $v_n$. A crucial property of repeater lines that enables the application of  $X$ protocol is that given a vertex $v_i$ its only neighbors that are part of the repeater line are $v_{i-1}$ and $v_{i+1}$. Note that this property is also satisfied by the shortest path connecting $v_1$ to $v_n$, and it is exactly the one used in proving the $X$ protocol in the original article (cf. Supplementary Information of Theorem 1 of \emph{Ref.} \cite{HPE19}). 

So far, we have only looked at how the edges among the repeater line vertices change with the $X$ protocol. The edges connecting the neighborhood vertices to the vertices along the repeater line will also update with each measurement. We need not keep track of those edges since anything connected to $v_1,v_n$ at the end will be deleted anyway. Since the $X$ protocol consists solely of local complementations and vertex deletions along the repeater line, the vertices that remain connected to $v_1$ and $v_n$ at the end of the protocol are a subset of the initial neighborhood of the repeater line. Hence, the total number of measurements required for the $X$ protocol is, at most, the ones required by the repeater protocol. This is the same argument used in \emph{Ref.} \cite{HPE19}, where they derive an expression for the total number of measurements (Supplementary Information, Equation 16). The expression coincides with our case, the only difference being that they used the shortest path, whereas any repeater line would suffice in ours.

  \begin{figure}
    \centering
    \includegraphics[width=
    \linewidth]{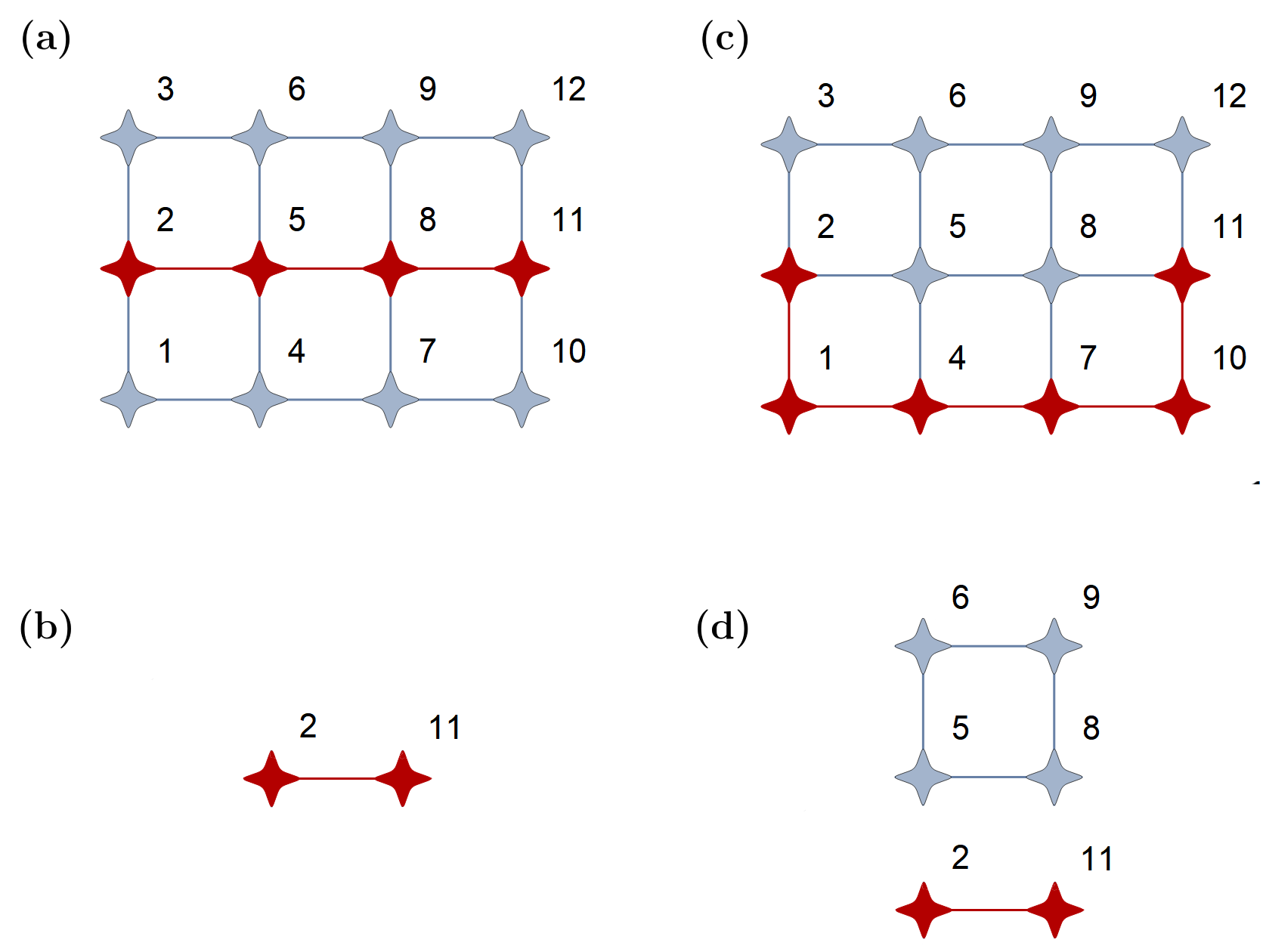}
    \caption{\textbf{(a)} Shortest path between $2,11$. \textbf{(b)}$X$ protocol performed along the shortest path just leaves us with the Bell pair shared between $2$ and $11$.  \textbf{(c)} A repeater line connecting $2,11$. \textbf{(d)} $X$ protocol along this path leaves much more connections in the graph.}
    \label{difcomb}
\end{figure}
\begin{figure}[b]
    \centering
    \includegraphics[width=
    \linewidth]{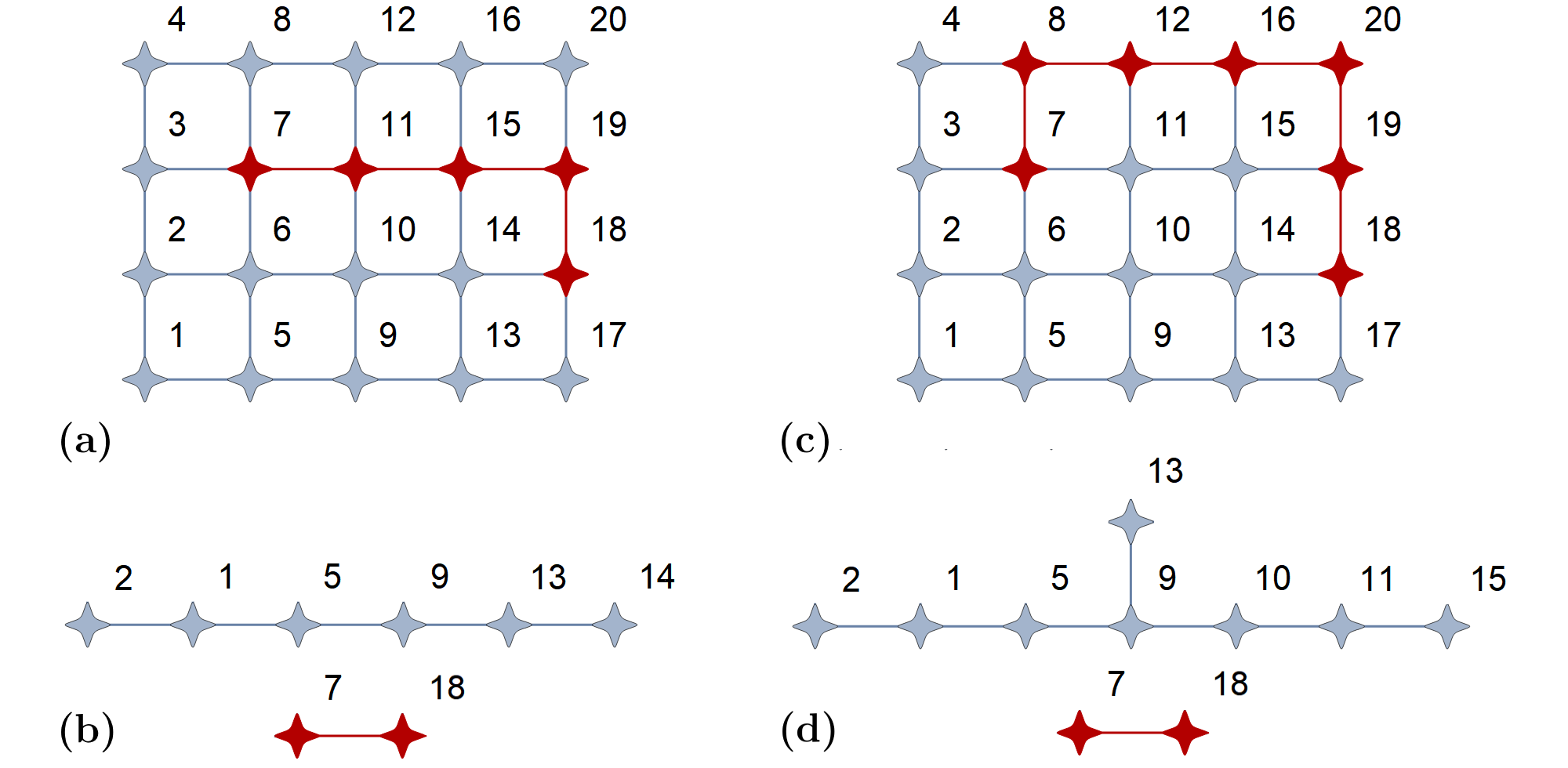}
    \caption{\textbf{(a)} Shortest path between $7,18$ with a combined neighborhood of $14$ vertices. \textbf{(b)} $X$ protocol along the shortest path. \textbf{(c)} A repeater line between $7,18$ with same amount of neighbourhood vertices. \textbf{(d)} $X$ protocol along the repeater line.}
    \label{samecomb}
\end{figure}

\autoref{difcomb} illustrates the $X$ protocol applied to the shortest path and a repeater line connecting the vertices $2,11$. $X$ protocol along the repeater line shown in \autoref{difcomb} (c) requires fewer measurements than along the shortest path shown in \autoref{difcomb} (a) and leaves more entanglement that can be utilized in subsequent communication rounds (compare \autoref{difcomb} (b) and (d)). Note that for the shortest path, there is no advantage in choosing the $X$ protocol over the repeater protocol, as both give the same result. However, the $X$ protocol along the repeater line outperforms the repeater protocol. This highlights the importance of our extension in exploiting the full potential of $X$ protocol and local complementation operations. Note that in this particular example, the neighborhood of the repeater line is smaller than that of the shortest path. In \autoref{samecomb}, we illustrate an example of the two paths with an equal amount of neighborhood vertices. In this particular case, even with the same number of neighborhood nodes, the repeater line proves to be a better alternative. Thus, considering repeater lines for performing $X$ protocol is generally better than restricting just to the shortest paths.\\
 \section{Bottlenecks}
 \label{sec:bottle}
The aforementioned routing schemes for entangling network nodes invoke the problem of overlapping paths when considering simultaneous communication requests from more than two pairs of nodes. \emph{Ref.} \cite{HPE19} shows how local complementation can help to solve such bottlenecks by demonstrating an example of the butterfly network, the smallest instance of the grid network. Recently, bottlenecks were also investigated in the context of nearest-neighbor architectures like ring and line networks \cite{HDE+22}. In \cite{HDE+22}, it was concluded that ring and line networks have severe limitations in overcoming bottlenecks, unlike the grid.
Here we point out that the butterfly network faces the same bottleneck issues as the other nearest-neighbor networks. Going beyond the butterfly, we also investigate bottlenecks in a similar grid network. We present an example of a butterfly-like network for which the same bottleneck problem cannot be overcome. The results demonstrate how even grid networks are not exempt from bottleneck concerns.

\begin{figure}[h]
    \centering
    \includegraphics[width=
    \linewidth]{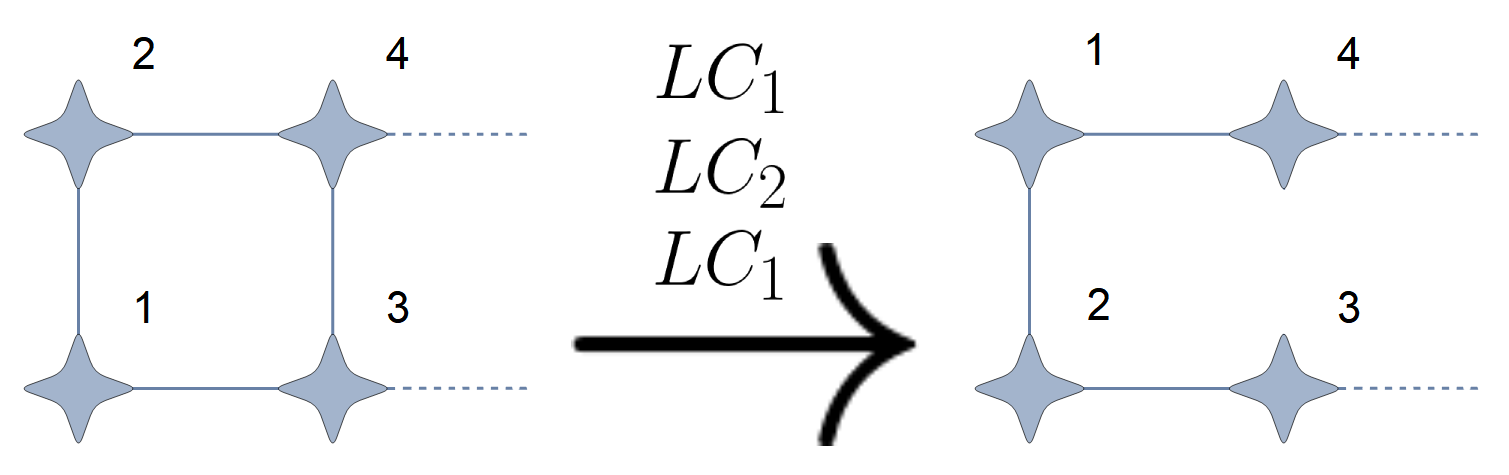}
    \caption{Transforming grid graphs to ring graphs via local complementations. Local complementation applied to vertices $1$, then on $2$, and then on $1$ again yields the graph on the right. The transformation removes the edge between $3,4$ and swaps $1$ and $2$. The rest of the graph (denoted by the dashed lines) is unaffected.}
    \label{lcc}
\end{figure}

 Before we discuss the results, it would be helpful to define an operation consisting of a set of local complementations on grid graphs. We denote this operation as $LC_{i,j}=LC_iLC_jLC_i$, where $LC_i$ is the local complementation on vertex $i$. The effect of this operation is depicted in \autoref{lcc}. If a graph can be converted to another using just a sequence of local measurements,  then the graphs are called \emph{locally equivalent}. We will utilize this to demonstrate the bottlenecks in grid graphs.
 
\begin{figure}
    \centering
    \includegraphics[width=
    \linewidth]{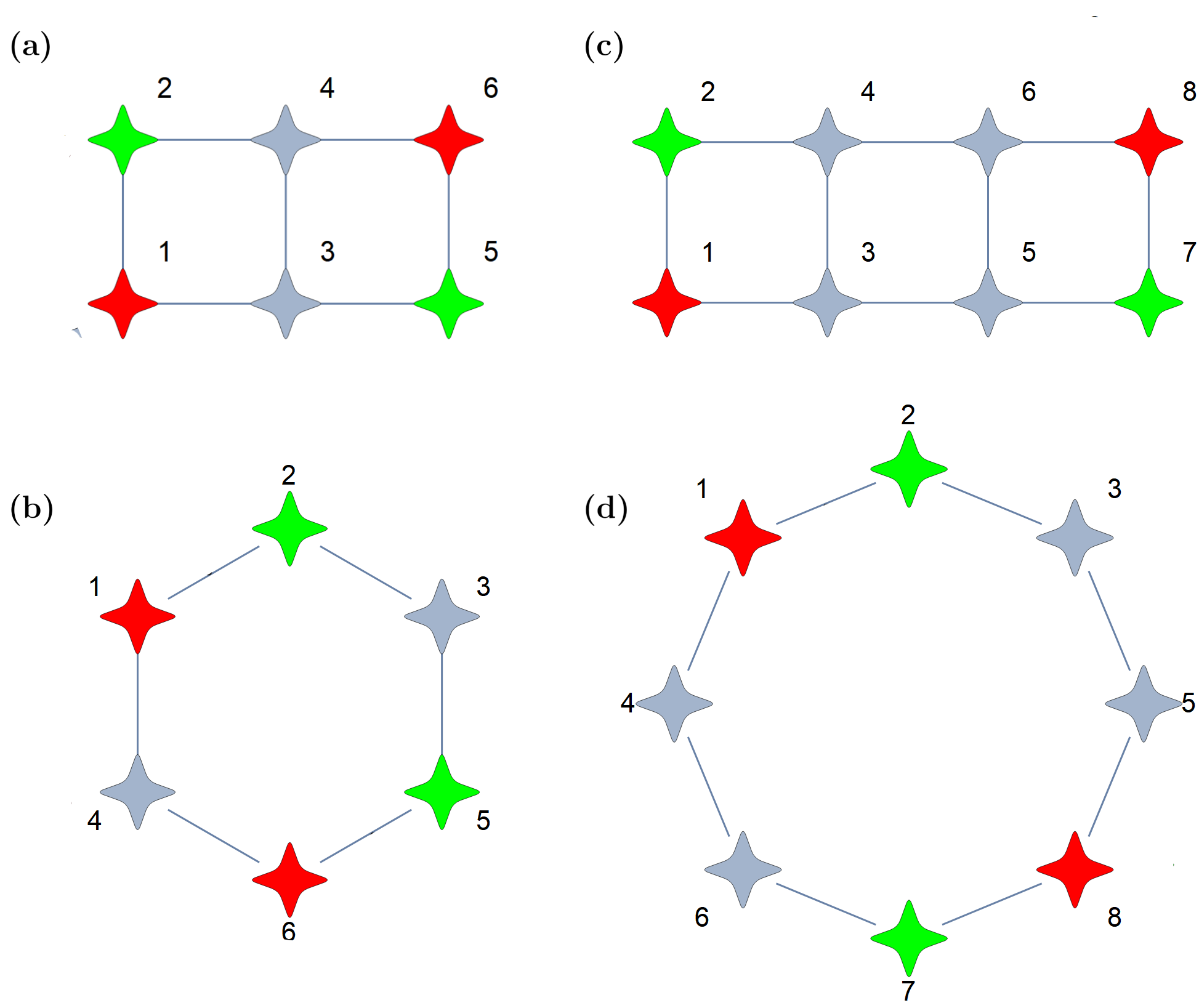}
    \caption{In the following, pairs of vertices with the same color (green/red) are to be connected. \textbf{(a)} The butterfly network. \textbf{(b)} The ring graph is locally equivalent to the butterfly network. Simultaneous pair generation is possible in this case using just local measurements. \textbf{(c)} $2\times4$ grid graph. \textbf{(d)} The ring graph locally equivalent to $2\times 4$ grid graph. Solving the bottleneck, in this case, is impossible since the paths connecting red-red and green-green always cross. }
    \label{butter}
\end{figure}

 The butterfly network is a special case of $2\times n$ grid network with $n=3$ (\autoref{butter}\textbf{(a)}). \emph{Ref.} \cite{HPE19} considered the bottleneck  of simultaneous Bell pair generation between $1,6$ and $2,5$. They showed that the problem could be solved using local complementations and vertex deletion or equivalently with  $X$ measurements on vertices $3,4$. This leaves us with two desired Bell pairs (Figure 2 of  \cite{HPE19}). However, this is only a particular example where local complementation helps with the bottleneck, and it is not possible in general. We use Theorem 1 of \cite{HDE+22} to establish this. In essence, the theorem states that extracting two Bell pairs from a ring is impossible if the paths connecting the pairs cross each other. For example, consider the pairs of vertices $1,3$ and $2,5$ in \autoref{butter}\textbf{(b)}. One can easily see that the paths connecting $1$ with $3$ and $2$ with $5$ overlap. Thus, it is impossible to extract the two Bell pairs from \autoref{butter}\textbf{(b)}. However, this ring graph is locally equivalent to the butterfly network \autoref{butter}\textbf{(a)}. Applying $LC_{1,2}$ converts the graph to one another. This implies that its impossible to extract entangled pairs among $1,3$ and $2,5$ in the butterfly network. Note that the case of $1,6$ and $2,5$ is different. As the paths between them do not cross in the locally equivalent ring graph, it is possible to overcome this bottleneck.
 A similar case for $2\times 4$ grid graph is depicted in \autoref{butter}\textbf{(c)}. \autoref{butter}\textbf{(d)} is its locally equivalent ring graph. Interestingly, the bottleneck between the two pairs of end nodes which were solvable in the butterfly network is impossible to solve in this case. In \autoref{butter}\textbf{(d)}, any routing path connecting $1,8$ and $2,7$ will have a common edge, a bottleneck. Using the same arguments as above, we can say that the bottleneck in \autoref{butter}\textbf{(c)}  is impossible to solve. Note that this technique of converting grid graphs to ring graphs can also be utilized in other cases. If such a conversion is possible, the no-go results of ring graphs can be applied directly to grid graphs.\\

\begin{figure}
      \centering
      \includegraphics[width=
    \linewidth]{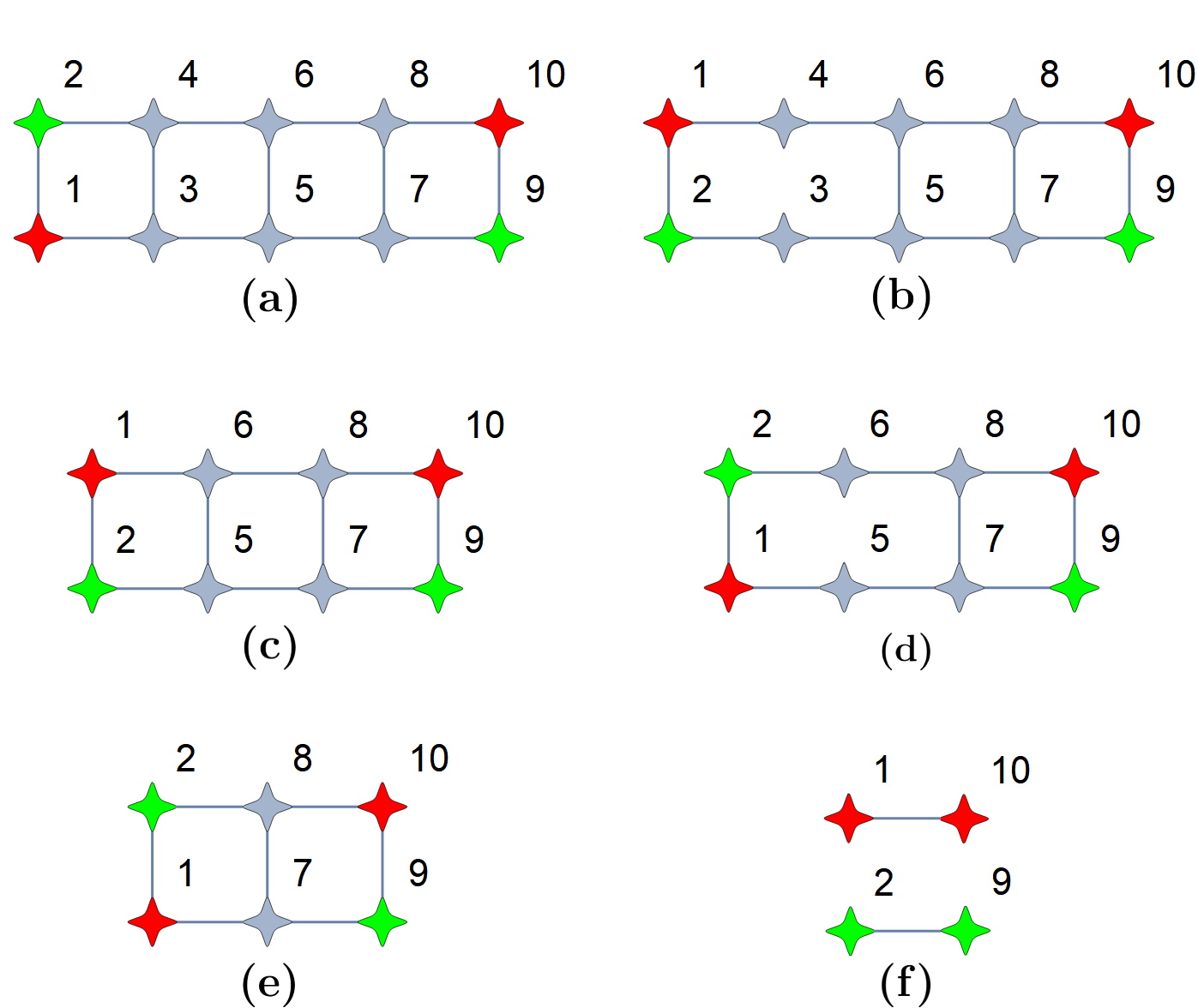}
      \caption{Bottlenecks between the same colored vertices in a $2\times 5$ grid network is solved using successive local complementation pairs and $Y$ measurements. \textbf{(a)}$2\times 5$ grid network. \textbf{(b)} Resulting graph after $LC_{1,2}$. \textbf{(c)} $Y$ measurements on $4$ and $3$ yields a $2 \times 4 $ grid graph. \textbf{(d)} After $LC_{1,2}$. \textbf{(e)}After $Y_6$ and $Y_5$. \textbf{(f)} After $X_8$ and $X_7$, the desired Bell pairs.  }
      \label{fig:2x5}
  \end{figure}

 In the general case of $ 2\times n $ graphs, when $n$ is odd, we see that the bottleneck between the vertices $1,2n$ and $2,2n-1$ can be solved. The trick is to perform successive $LC_{1,2}$ operations and $Y$ measurements (proposition \autoref{prop:ymeas}) on the now disconnected vertices until we reach the butterfly network from where it can be solved using $X$ measurements.  Local corrections are required on the qubits after every  $Y$ measurement (\cite{HEB04}). In  \autoref{fig:2x5}, an example for the case $2\times 5$ is detailed for clarity.   However, this technique does not apply to the case when $n$ is even since in that case, we end up with a similar  bottleneck as in the $2\times 4 $ case, which cannot be solved (\autoref{butter}). 
 \section{Discussions}
 \label{sec:discussion}
We have expanded upon the original $X$ protocol of \emph{Ref.} \cite{HPE19} to incorporate more than the shortest paths. A general path's properties were defined to work with the $X$ protocol. We have provided examples where the shortest paths yielded sub-optimal results compared to our generalized version. We also discussed bottlenecks in grid graphs. Utilizing a graph transformation operation on grid graphs, we showed that grid graphs suffer from the same bottleneck issues as other nearest-neighbor networks, such as ring graphs. An interesting future research direction would be to explore the utility of the graph theoretic results of \cite{HDE+22} in proving similar no-go results for grid graphs. It would also be of interest to explore bottlenecks in generating simultaneous multiparty entangled states \cite{MP22}. Identifying a repeater line connecting the nodes is necessary for this protocol. Hence it will face the same issues discussed here while dealing with simultaneous communication requests.

The results discussed here would be helpful in exploring instances of measurement-based quantum network coding \cite{MSN+18}. In quantum network coding, multiple qubits are transmitted simultaneously through the network, and the information is encoded in the entangled state of the qubits. This allows for a more efficient transmission of information, as multiple pieces of data can be sent in a single transmission. Ref. \cite{MSN+18} explored fidelity of communication in butterfly networks, while using the method outlined in the previous section to overcome the communication bottlenecks. In the specific case they considered, it improved the noise tolerance compared to the traditional methods based on entanglement swapping. Our results on resolving bottlenecks in more general network structures clearly reveal interesting possibilities for future research on network coding. We hypothesize that improvements from such measurement-based schemes over the traditional ones will increase as the number of network nodes/connections increases.

\section*{Acknowledgment}
VM and AP acknowledge support from the QUEST scheme of the Interdisciplinary Cyber-Physical Systems (ICPS) program of the Department of Science and Technology (DST), India, Grant No.: DST/ICPS/QuST/Theme-1/2019/14 (Q80).

\bibliographystyle{naturemag}
\bibliography{network.bib}

\end{document}